\documentclass[prc,twocolumn,superscriptaddress,showpacs]{revtex4}
\usepackage{graphicx,amsmath,amssymb,bm,multirow}

\newcommand{\be}{\begin{equation}}
\newcommand{\ee}{\end{equation}}
\newcommand{\bea}{\begin{eqnarray}}
\newcommand{\eea}{\end{eqnarray}}

\newcommand{\kf}{k_{\rm F}}

\newcommand{\pp}{{\bf p}}
\newcommand{\kk}{{\bf k}}
\newcommand{\PP}{{\bf P}}

\newcommand{\rme}{{\rm e}}
\newcommand{\rmi}{{\rm i}}

\begin{document}

\title{Spin response of a normal Fermi liquid with noncentral interactions}

\author{C.\ J.\ Pethick}
\email[E-mail:~]{pethick@nbi.dk}
\affiliation{The Niels Bohr International Academy, The Niels Bohr Institute,
Blegdamsvej 17, DK-2100 Copenhagen \O, Denmark}
\affiliation{NORDITA, Roslagstullsbacken 21, 10691 Stockholm, Sweden}
\author{A.~Schwenk}
\email[E-mail:~]{schwenk@triumf.ca}
\affiliation{TRIUMF, 4004 Wesbrook Mall, Vancouver, BC, V6T 2A3, Canada}


\begin{abstract}
We consider the spin response of a normal Fermi liquid with
noncentral interactions under conditions intermediate between the
collisionless and hydrodynamic regimes.  This problem is of
importance for calculations of neutrino properties in dense matter.
By expressing the deviation of the quasiparticle distribution
function from equilibrium in terms of eigenfunctions of the
transport equation under the combined influence of collisions and an
external field, we derive a closed expression for the
spin-density--spin-density response function and compare its
predictions with that of a relaxation time approximation. Our results
indicate that the relaxation time approximation is reliable for 
neutrino properties under astrophysically relevant conditions.
\end{abstract}

 
\maketitle

\section{Introduction}

Neutrino scattering, production, and annihilation rates are key
physical quantities in understanding stellar collapse, neutron star
formation, and supernovae, and they are determined by the spin,
isospin and density responses of nuclear matter~\cite{raffelt}.  The
challenge is to include the effects of collisions between nucleons and
mean-field effects in a consistent way.  This problem has been
approached in the framework of Landau's theory of normal Fermi
liquids~\cite{olsson} but, in the detailed calculations performed to
date, the collision integral has been replaced by a simple expression
with a single relaxation time~\cite{lykasov}.

In this paper we consider the spin response at long wavelengths.  This
problem is well understood in two limiting cases.  The first is the
collisionless regime, where the collision rate is small compared with
the frequency~$\omega$ of the field, and collisional effects may be
included perturbatively~\cite{olsson}.  When collisions cannot be
taken into account perturbatively, they give rise to the
Landau--Pomeranchuk--Migdal effect, which in the language of
perturbation theory is due to energy denominators acquiring an
imaginary part due to collisional damping of excitations~\cite{LPM}.

The second regime that is well understood is the hydrodynamic one,
where the collision rate is large compared with the frequency.  In
this case, the transport equation for quasiparticles may be solved
exactly using the methods developed by Jensen, Smith and
Wilkins~\cite{jensen} and by Brooker and Sykes~\cite{bs}.  In
astrophysical applications, conditions intermediate between these two
limiting cases arise.  Related problems for the spin-independent part
of the response, corresponding to distortions of the quasiparticle
distribution in momentum space with an angular dependence proportional
to a spherical harmonic, have previously been attacked using
variational methods.  These were developed in the context of the
frequency-dependent electrical conductivity and magnetoresistance of a
normal metal~\cite{Ah-Sam}, and provide upper and lower bounds for the
imaginary part of the response.  These methods were subsequently
exploited to obtain upper and lower bounds on the response of the
quadrupolar distortion of the quasiparticle distribution, which is a
key feature in calculations of the attenuation of sound in the regime
intermediate between the first and zero sound~\cite{egilsson}.

Exact expressions for relaxational modes of a Fermi liquid have
previously been derived in the regime intermediate between the
collisionless and hydrodynamic ones in Ref.~\cite{bs1972}. In that
paper, eigenfunctions of the Fourier transform (with respect to time)
of the Boltzmann transport equation for quasiparticles were found.

In the present paper we extend the work on exact solutions to include
an external field, in contrast to most earlier work with exact
solutions which considered transport coefficients in the hydrodynamic
limit, where the effects of the external field drop out
\cite{jensen,bs}, or involved the spectra of relaxational modes and
collective modes in the absence of an external field \cite{bs1972}.
We use the eigenfunctions obtained in Ref.\ \cite{bs1972} to derive a
closed expression for the spin-density--spin-density response valid
generally.  We compare our results with those of a single relaxation
time approximation, with a relaxation time chosen to reproduce the
correct behavior in the collisionless regime, and find that for
astrophysically relevant conditions, the response function given by
the relaxation time approximation differs from the exact result by
less than $10\%$.

\section{Basic formalism}

We calculate the response of a normal Fermi liquid to a magnetic field
in the $z$-direction that is spatially homogeneous and varies in time
as $\rme^{-\rmi \omega t}$.  We assume that the quasiparticle density
matrix and the quasiparticle energy may be taken to be diagonal in
$\sigma_z$.  In doing so, we neglect the fact that noncentral
components of the interaction, which are very important in systems of
nucleons, can create nonzero $x$- and $y$-components of the
spin-dependent part of the quasiparticle density matrix.  In addition,
we neglect tensor components of the magnetic moment.  In the presence
of a uniform magnetic field, these would give rise to a quadrupolar
distortion of the quasiparticle distribution as a function of
direction in momentum space~\cite{dabrowski}.  Therefore we write the
change in the energy $\delta \varepsilon_{\pp \sigma}$ of a
quasiparticle with momentum $\pp$ and $z$-component of the spin
($\sigma=\pm 1$) due to the application of the external field $U_z$ as
\be
\delta \varepsilon_{\pp \sigma} = \sigma \, U_z \,,
\label{extfield}
\ee
and denote the quasiparticle distribution function as $n_{\pp \sigma}$.

We start from the quasiparticle kinetic equation, which for
spatially uniform conditions is given by
\be
\frac{\partial n_{\pp \sigma}}{\partial t} = I[n_{\pp' \sigma'}] \,,
\label{transport}
\ee
where $I[n_{\pp' \sigma'}]$ is the collision integral.  In normal
Fermi liquids at low temperatures, the dominant collisions are those
between pairs of quasiparticles and, in a compact notation in which
$\pp_i \sigma_i$ is denoted simply by $i$, Eq.~(\ref{transport})
takes the form (see Ref.~\cite{BaymPethick} for details)
\begin{align}
\frac{\partial n_1}{\partial t} &= - n_1 \, 2 \pi \sum_{234} \frac{1}{2}
\bigl|\langle12|{\cal A}|34\rangle\bigr|^2 n_2 \, (1-n_3) \, (1-n_4) \,
\delta_\varepsilon \delta_\pp \nonumber \\
&+ (1-n_1) \, 2 \pi \sum_{234} \frac{1}{2}
\bigl|\langle34|{\cal A}|12\rangle\bigr|^2
(1-n_2) \, n_3 \, n_4 \, \delta_\varepsilon \delta_\pp \,,
\label{kineticeq}
\end{align}
where $\langle12|{\cal A}|34\rangle$ is the quasiparticle scattering
amplitude, $\delta_\varepsilon \delta_\pp$ is shorthand for
$\delta(\varepsilon_1
+\varepsilon_2-\varepsilon_3-\varepsilon_4)\delta(\pp_1+\pp_2-\pp_3-\pp_4)$
and $\sum_i = \sum_{\sigma_i} \int d \pp_i/(2\pi)^3$. The two
quasiparticles in the scattering process are of the same species and
the factor $1/2$ avoids double counting of final states when momenta
are freely summed over.
In addition, we assume that $\omega \ll T$, so we do not include
$\omega$ in the energy-conserving delta function~\cite{units}.

The quasiparticle energies include contributions due to the external
field and due to ``molecular fields'' resulting from interactions with
other quasiparticles,
\be
\varepsilon_{\pp \sigma} =
\varepsilon^0_{\pp \sigma} + \delta \varepsilon_{\pp \sigma} \,,
\ee
where $\varepsilon_{\pp \sigma}^0$ denotes the quasiparticle energy
in equilibrium in the absence of the magnetic field and $\delta 
\varepsilon_{\pp \sigma}$ includes the interaction with the magnetic
field, Eq.~(\ref{extfield}), and the contribution due to 
quasiparticle interactions $f_{12}$,
\be
\delta \varepsilon_1 = \sigma_1 \, U_z +\sum_2 f_{12} \, \delta n_2 \,.
\ee
Tensor components of the Landau quasiparticle interaction are
generally small in nuclear matter and we therefore neglect them. In
this case the response of the quasiparticle distribution to a
spatially uniform magnetic field is isotropic and the interaction
energy reduces to $g_0 S$, where
$S = \sum_1 \sigma_1 \, \delta n_1$ is the total spin and $g_0$
(or $f_0^a$ in the quantum liquids' literature~\cite{Pines}) is the
isotropic component of the spin-dependent interaction between
quasiparticles. The change in the quasiparticle energy is thus
\be
\delta
\varepsilon_1 = \sigma_1 (U_z + g_0 S) \,.
\ee

We now linearize Eq.~(\ref{kineticeq}) about the equilibrium distribution
function
\be
n_i^0 = \frac{1}{\rme^{(\varepsilon_i^0-\mu)/T}+1} \,,
\ee
in the standard way~\cite[Sect.~1.2.4]{BaymPethick} taking into account
the variation of the quasiparticle energies in the delta function.
Here $\mu$ is the chemical potential in equilibrium.  It is convenient
to introduce the quantity
\be
\delta \overline{n}_i = n_i - n^0(\varepsilon_i)
= n_i - n_i^0 - \frac{\partial n_i^0}{\partial \varepsilon^0_i}
\delta \varepsilon_i \,,
\ee
which represents the difference between the quasiparticle distribution
function and the Fermi function evaluated for the actual quasiparticle
energy, not the one in equilibrium. Linearizing Eq.~(\ref{kineticeq})
leads to
\begin{widetext}
\vspace*{-4mm}
\begin{align}
\frac{\partial \delta \overline{n}_1}{\partial t} +
\sigma_1 \frac{\partial n^0_1}{\partial \varepsilon^0_1}
\frac{\partial (U_z +g_0 S)}{\partial t}
&= - \frac{2\pi}{T} \, \sum_{234} \, \frac{1}{2} \, \bigl|\langle12|{\cal A}|34
\rangle\bigr|^2 \, \delta_\pp \, \delta_x \, 
\biggl( \, \prod\limits_{i=1}^4 \frac{1}{2 \cosh(x_i/2)} \biggr)
\nonumber \\[1mm]
&\times \bigl[2 \cosh(x_1/2) \, \Xi_1+2 \cosh(x_2/2) \, \Xi_2
-2 \cosh(x_3/2) \, \Xi_3 -2 \cosh(x_4/2) \, \Xi_4 \bigr] \,,
\label{kineticeq2}
\end{align}
\end{widetext}
where $x_i = (\varepsilon_i^0-\mu)/T$, so that one has $(\partial
n^0_i/ \partial \varepsilon^0_i)^{-1} = - T \, [2 \cosh(x_i/2)]^2$,
and $\delta_x = \delta(x_1+x_2-x_3-x_4)$. In addition, we
have introduced $\Xi_i = 2 \cosh(x_i/2) \, \delta \overline{n}_i$,
which simplifies the subsequent calculations. In writing the
right-hand side of Eq.~(\ref{kineticeq2}) in this form,
we have used the property
that for interactions invariant under time reversal $\langle \pp_1 \sigma_1,
\pp_2 \sigma_2 |{\cal A}| \pp_3 \sigma_3, \pp_4 \sigma_4 \rangle =
(\langle -\pp_3 -\sigma_3, -\pp_4 -\sigma_4 |{\cal A}| -\pp_1
-\sigma_1, -\pp_2 -\sigma_2\rangle)^*$, and therefore, since the
distribution function is isotropic in momentum space, one may
combine the two terms in Eq.~(\ref{kineticeq}). After Fourier 
transforming with respect to time, Eq.~(\ref{kineticeq2}) may be
recast as an inhomogeneous integral equation for $\Xi$,
\begin{widetext}
\vspace*{-6mm}
\begin{align}
\frac{\sigma_1 (U_z +g_0 S)}{2 T \cosh(x_1/2)}
&= \Xi_1 + \frac{\rmi}{\omega} 
\frac{2\pi}{T} \, \sum_{234} \, \frac{1}{2} \, \bigl|\langle12|{\cal A}|34
\rangle\bigr|^2 \, \delta_\pp \, \delta_x \, 
\biggl( \, \prod\limits_{i=2}^4 \frac{1}{2 \cosh(x_i/2)} \biggr)
\nonumber \\[1mm]
&\times \bigl[2 \cosh(x_1/2) \, \Xi_1+2 \cosh(x_2/2) \, \Xi_2
-2 \cosh(x_3/2) \, \Xi_3 -2 \cosh(x_4/2) \, \Xi_4 \bigr] \,.
\label{KXi}
\end{align}
\end{widetext}
For $T \ll \varepsilon_{\rm F}$, scattering processes are strongest
for quasiparticle states in the vicinity of the Fermi surface. As
a consequence, integrals over the magnitudes of the momenta may be
decoupled from integrals over quasiparticle energies, and the
scattering amplitudes involved can be replaced by their values
for all quasiparticle momenta on the Fermi surface. In the
literature on quantum liquids it is customary to write~\cite{BaymPethick,AK}
\be
d{\pp}_2 = p_2^2 \, dp_2 \, d\cos\theta \, d\phi_2 \,,
\ee
where $\theta$ is the angle between $\pp_1$ and $\pp_2$ and $\phi_2$
is an azimuthal angle, and
\be
d{\pp}_3 = p_3^2 \, dp_3 \, d\cos\theta_3 \, d\phi \,,
\ee 
where $\theta_3$ is the angle between $\pp_3$ and the total momentum
$\PP=\pp_1+\pp_2$ of the incoming quasiparticles and $\phi$ is the
angle between the plane containing $\pp_1$ and $\pp_2$ and that
containing $\pp_3$ and $\pp_4$. The integral over $\theta_3$ can be
converted into one over $\varepsilon_4$ by using the fact that $\pp_4
=\PP -\pp_3$, and therefore
\be
p_4^2 = P^2 + p_3^2 - 2 P p_3 \cos \theta_3 \,,
\ee
from which is follows that
\be
p_4 dp_4 = -P p_3 \, d\cos\theta_3 \,,
\ee
when $p_3$ is held fixed, or
\be
\biggl| \frac{\partial \cos\theta_3}{\partial \varepsilon_4} \biggr|
= \frac{m^*}{P \kf} \quad \text{for fixed $p_3$} \,.
\ee
Thus we may write
\be
d{\pp}_2 \, d{\pp}_3 = \frac{m^{*3} \kf}{P} \, d\varepsilon_2 \,
d\varepsilon_3 \, d\varepsilon_4 \, d\cos\theta \, d\phi \, d\phi_2 \,,
\ee
where $P=2 \kf \cos(\theta/2)$.

Next we express $\Xi$ as
\be
\Xi_i = \sigma_i \, \frac{(U_z +g_0S)}{T} \, \xi(x_i) \,,
\label{Xi}
\ee
substitute this expression into Eq.~(\ref{KXi}), multiply the result
by $\sigma_1$, average over $\sigma_1$, and perform the integrals. 
The integral over $\phi_2$ gives $2\pi$ and the term involving $\Xi_1$
is proportional to the integral $K_3$, Eq.\ (\ref{K3}). One thus
obtains the linearized quasiparticle transport equation,
\begin{widetext}
\vspace*{-5mm}
\be
\frac{1}{2 \cosh(x_1/2)} = \frac{\rmi}{2\omega\tau_0} \, ( x_1^2
+ \pi^2 \zeta^2 ) \, \xi(x_1) - \frac{\rmi \lambda_\sigma}{
\omega \tau_0} \int_{-\infty}^\infty \frac{dx_2 dx_3 dx_4 \,
\delta_x}{4 \cosh(x_2/2) \cosh(x_4/2)} \, \xi(x_3) \,.
\label{Kxi}
\ee
\end{widetext}
Here we have introduced the quantities
\begin{align}
\zeta^2 &= 1 - \rmi \, \frac{2 \omega \tau_0}{\pi^2} \,, \\[1mm]
\frac{1}{\tau_0} &= \frac{\pi}{16} \frac{T^2}{\kf v_{\rm F}} \,
\biggl\langle {\rm Tr}_\sigma \bigl|\langle12|{\cal A}|34
\rangle\bigr|^2 \biggl\rangle \,, \label{tau0} \\[1mm]
\lambda_\sigma &= \frac{\biggl\langle {\rm Tr}_\sigma 
\bigl|\langle12|{\cal A}|34
\rangle\bigr|^2 \, \sigma_1(-\sigma_2+\sigma_3+\sigma_4)
\biggl\rangle}{\biggl\langle {\rm Tr}_\sigma \bigl|\langle12|{\cal A}|34 
\rangle\bigr|^2 \biggl\rangle} \,,
\label{lambda}
\end{align}
where the trace ${\rm Tr}_\sigma$ denotes a sum over all spins
$\sigma_i$ and the average is defined by
\be
\bigl\langle \, \ldots \bigl\rangle \, = \frac1{8\pi} \,
\int_0^\pi \frac{\sin \theta \, d\theta}{\cos(\theta/2)} \int_0^{2\pi}
d\phi \: \ldots \,,
\label{angav}
\ee
and $v_{\rm F}=\kf/m^*$ is the Fermi velocity. The factor of
$1/(8\pi)$ ensures that the average is normalized to 
unity~\cite{normalization}.
The first term on the right-hand side of Eq.~(\ref{Kxi}) includes the
contribution from the integral over $x_2$, $x_3$ and $x_4$ given by
the function $K_3$, Eq.~(\ref{K3}), in Appendix~\ref{math}.  The
integrals over $x_3$ and $x_4$ in Eq.~(\ref{Kxi}) yield the standard
form for the kernel, given by the function $K_2$, Eq.~(\ref{K2}).
However, for the Fourier transformation in the next section, it is
simpler if the integral is left in the original form.

Physically, the quantity $2\tau_0/\pi^2$ is the relaxation time of a
quasiparticle in a particular momentum state at the Fermi surface, if
deviations of the quasiparticle distribution from equilibrium in all
other states vanish. For an interaction with only central terms, the
expression, Eq.~(\ref{tau0}), reduces to that derived in the context
of liquid $^3$He~\cite{notation}. 
The quantity $1-\lambda_\sigma$ is a measure of
the effectiveness of collisions in relaxing the spin of the
quasiparticle: if spin is conserved in collisions,
$\sigma_1+\sigma_2=\sigma_3+\sigma_4$ and therefore
$\lambda_\sigma=1$, while if $\lambda_\sigma=0$, the integral term in
Eq.~(\ref{Kxi}) vanishes and the equation becomes an algebraic one,
whose physical content is equivalent to a relaxation time
approximation with a relaxation time that depends on the quasiparticle
energy, as we will see in Sect.~\ref{examples}.

In applications it is sometimes convenient to express the scattering
amplitudes in terms of the momentum transfers $\kk=\pp_1-\pp_3$ and
$\kk'=\pp_1-\pp_4$.  For quasiparticles at the Fermi surface these
satisfy the condition 
\be
P^2+k^2+k'^2=4\kf^2 \,.
\ee
The magnitudes of the momentum transfers are given by $k=2\kf 
\sin(\theta/2) \sin(\phi/2)$ and $k'=2\kf \sin(\theta/2) \cos(\phi/2)$
and the Jacobian for the transformation is
\be
\biggl| \frac{\partial(\cos\theta,\phi)}{\partial(k,k')} \biggr| 
= \frac{2}{\kf^2} \,,
\ee
and therefore~\cite{lykasov}
\be
\bigl \langle \ldots \bigr \rangle \, = \frac{1}{\pi}
\int_0^{2 \kf} \frac{dk}{\kf} \, \int_0^{2 \kf} \frac{dk'}{\kf} \: 
\frac{\kf \, \Theta(4 \kf^2 - k^2 - k^{\prime\,2})}{
\sqrt{4 \kf^2 - k^2 - k^{\prime\,2}}} \: \ldots \,.
\ee
When effects of the medium on scattering amplitudes are taken into
account, the scattering amplitude will also depend on the total
momentum of the incoming quasiparticles, and the natural variables
for scattering of quasiparticles at the Fermi surface are $P$ and
$\phi$. In terms of these the average, Eq.~(\ref{angav}), takes on
the simple form
\be
\bigl \langle \ldots \bigr \rangle \, = \int_0^{2 \kf}
\frac{dP}{2\kf} \, \int_0^{2 \pi} \frac{d\phi}{2\pi}
\: \ldots \,.
\ee
 
In the next section, we solve Eq.~(\ref{Kxi}) for $\xi$, from which
one can calculate the total spin $S$,
\begin{align}
S &= \sum_1 \sigma_1 \, \delta n_1 = \sum_1 \sigma_1
\biggl( \delta \overline{n}_1 + \sigma_1 \frac{\partial n_1^0}{\partial
\varepsilon^0_1} \, (U_z +g_0 S) \biggr) \,, \nonumber \\
&= \sum_1 \sigma_1 \biggl( \frac{\Xi_1}{2\cosh (x_1/2)} + \sigma_1
\frac{\partial n_1^0}{\partial \varepsilon^0_1} \, (U_z +g_0 S) \biggr) \,.
\label{totalspin}
\end{align}
Since $\Xi$ is proportional to $U_z +g_0 S$, this equation shows that
$S$ can be written in the form
\be
S = - \chi_\sigma U_z \,,
\ee
where the spin-density--spin-density response function $\chi_\sigma$
is given by
\be
\chi_\sigma = \frac{X_\sigma}{1+g_0 X_\sigma} \,,
\ee
and $X_\sigma = - S/(U_z +g_0 S)$ denotes the response function in the 
absence of mean-field effects ($g_0 = 0$). Using Eq.~(\ref{Xi}), it
follows from Eq.~(\ref{totalspin}) that
\be
X_\sigma = N(0) \biggl(1 -\int_{-\infty}^\infty \frac{dx}{2 \cosh(
x/2)} \, \xi(x) \biggr) \,,
\label{chi0}
\ee 
with $N(0) = - \sum_1 \partial n_1^0/\partial \varepsilon^0_1$
being the density of states at the Fermi surface.

\section{Solution of transport equation}

To solve the linearized transport equation, Eq.~(\ref{Kxi}), we 
follow the methods developed by Brooker and Sykes~\cite{bs1972} and
express the solution $\xi$ in the form
\be
\xi(x) = \sum_{r=0}^\infty a_r \, \Phi_r(x) \,,
\label{expansion}
\ee
where $\Phi_r(x)$ are eigenfunctions of the homogeneous equation,
\begin{multline}
( x_1^2 + \pi^2 \zeta^2 ) \, \Phi_r(x_1) \\[1mm]
= \frac{\Lambda_r}{2} 
\int_{-\infty}^\infty \frac{dx_2 dx_3 dx_4 \,
\delta_x}{\cosh(x_2/2) \cosh(x_4/2)} \, \Phi_r(x_3) \,.
\label{eigenfunction}
\end{multline}
The eigenfunctions may be determined by Fourier transforming. We write
\be
\widetilde{\Phi}_r(k) = \int_{-\infty}^\infty dx \: \Phi_r(x) \,
\rme^{-\rmi kx} \,.
\ee
By using the representation, Eq.~(\ref{delta}), for the delta
function, one finds that the Fourier transform of the integral
term reduces to the product of $ \widetilde{\Phi}_r(k) $ and
the square of the Fourier transform of $1/\cosh(x/2)$, given by
$2\pi/\cosh(\pi k)$. Thus the eigenfunctions satisfy
\be
-\frac{d^2 \widetilde{\Phi}_r(k)}{dk^2} + \pi^2 \zeta^2 \, \widetilde{
\Phi}_r(k) - \frac{2 \pi^2 \Lambda_r}{\cosh^2(\pi k)} \, \widetilde{
\Phi}_r(k) = 0 \,,
\label{Phi_r}
\ee
which has the same form as the Schr\"odinger equation for
a particle moving in a
one-dimensional potential $\sim {\rm sech}^2(\pi k)$ with a
(generally complex) energy $\sim \zeta^2$~\cite{LL}.  This is analogous to
the Sturmian method in quantum-mechanical problems, in which one uses
a basis of states with the same energy, but with different strengths
of the potential~\cite{avery}.

In this paper we are concerned with disturbances that are even in
$x$, and therefore the relevant eigenvalues are
\be
\Lambda_r = \frac{1}{2} \, (2r+\zeta)(2r+1+\zeta) \,,
\ee
where $r$ is positive integer or zero, and the corresponding
solutions~\cite[p.~85]{bs1972} are given by
\begin{align}
\widetilde{\Phi}_r(k) &= 2^\zeta \, \Gamma(1+\zeta) \, 
P_{2r+\zeta}^{-\zeta}\bigl(\tanh(\pi k)\bigr) \,, \\[1mm]
&= \frac{(2r)! \, \Gamma(2\zeta + 1)}{\Gamma(2r + 2\zeta + 1)} \,
[\cosh(\pi k)]^{-\zeta} \,
C_{2r}^{\zeta+1/2}\hspace*{-0.5mm}\bigl(\tanh(\pi k)\bigr) ,
\label{gegenbauer} \\[1mm]
&= {_2}F_1\bigl(-2r, 2r + 1 + 2\zeta, 1 + \zeta, [1 - \tanh(\pi
k)]/2 \bigr) \nonumber \\
& \quad \times [\cosh(\pi k)]^{-\zeta} \,,
\end{align}
where $C_n^m$ are Gegenbauer polynomials and ${_2}F_1$ is the
hypergeometric function.  The corresponding eigenfunctions as a
function of energy, $\Phi_r(x)$, may be determined by Fourier
transformation, and some of their properties are described in
Appendix~\ref{math}.

On substituting Eq.~(\ref{expansion}) into Eq.~(\ref{Kxi}),
multiplying by $\Phi_r(x_1)$ and integrating over $x_1$ 
one finds
\be
a_r = - \rmi 2 \omega \tau_0 \,
\frac{T_r}{U_r (\Lambda_r-\lambda_\sigma)} \,,
\ee
where, following the notation of Ref.~\cite{bs1972}, we have
defined
\be
T_r = \int_{-\infty}^\infty \frac{dk \, \widetilde{\Phi}_r(k)}{\cosh
(\pi k)} = \frac{\Gamma(\zeta +1) \, \Gamma(\frac{1}{2}+r+\frac{1}{2}
\zeta) \, \Gamma(\frac{1}{2}+r)}{\pi \, \Gamma(1+r+\frac{1}{2}\zeta)
\, \Gamma(1+r+\zeta)} \,,
\ee
and the normalizing integral is given by
\be
U_r = 2\pi \hspace*{-1mm} \int_{-\infty}^\infty 
\frac{dk \, \bigl[ \widetilde{\Phi}_r(k) \bigr]^2}{\cosh^2(\pi k)} =
\frac{2^{2\zeta+1} (2r)! \, [\Gamma(\zeta+1)]^2}{(2r+\zeta+\frac12) \,
\Gamma(2r+2\zeta+1)} .
\label{Ur}
\ee
The integral in Eq.~(\ref{chi0}) therefore leads to
\begin{align}
\int_{-\infty}^\infty \frac{dx \, \xi(x)}{2 \cosh(x/2)}
&= \sum_{r=0}^\infty \, \frac{a_r T_r}{2} \,, \nonumber \\[1mm]
&= - \rmi \omega \tau_0 \,
\sum_{r=0}^\infty \, \frac{(T_r)^2}{U_r} \, \frac{1}{\Lambda_r-
\lambda_\sigma} \,,
\end{align}
and the response function $X_\sigma$ in the absence of mean-field
effects is given by
\begin{widetext}
\vspace*{-5mm}
\begin{align}
\frac{X_\sigma}{N(0)} &= 1 + \rmi \, \frac{\omega\tau_0}{\pi^2} \,
\sum_{r=0}^\infty \, \frac{(2r+\zeta+\frac{1}{2}) \,
\Gamma(2r+2\zeta+1)}{2^{2\zeta +1} (2r)! \,
(\Lambda_r-\lambda_\sigma)} \biggl(
\frac{\Gamma(\frac{1}{2}+r+\frac{1}{2}\zeta) \,
\Gamma(\frac{1}{2}+r)}{
\Gamma(1+r+\frac{1}{2}\zeta) \, \Gamma(1+r+\zeta)} \biggr)^2 \,,
\nonumber \\[1mm]
&= 1 + \rmi \, \frac{2\omega\tau_\sigma(1-\lambda_\sigma)}{3} \,
\sum_{r=0}^\infty \, \frac{(2r+\zeta+\frac{1}{2}) \,
\Gamma(2r+2\zeta+1)}{2^{2\zeta +1} (2r)! \,
(\Lambda_r-\lambda_\sigma)} \biggl(
\frac{\Gamma(\frac{1}{2}+r+\frac{1}{2}\zeta) \,
\Gamma(\frac{1}{2}+r)}{
\Gamma(1+r+\frac{1}{2}\zeta) \, \Gamma(1+r+\zeta)} \biggr)^2 \,.
\label{chi0final}
\end{align}
\end{widetext}
The sum in Eq.~(\ref{chi0final}) may be expressed in terms of the
generalized hypergeometric function ${_5}F_4$ but in applications
it offers no advantages compared with a direct evaluation of the sum.
Moreover, we write the result in terms of the physically
relevant timescale in spin relaxation, $\tau_\sigma$, which for the
case of interest here, $\omega\ll T$, is given by
\be
\tau_\sigma = \frac{3 \tau_0}{2 \pi^2 (1-\lambda_\sigma)} \,.
\label{tau_sigma}
\ee

\begin{figure*}[t]
\begin{center}
\includegraphics[clip=,scale=0.48]{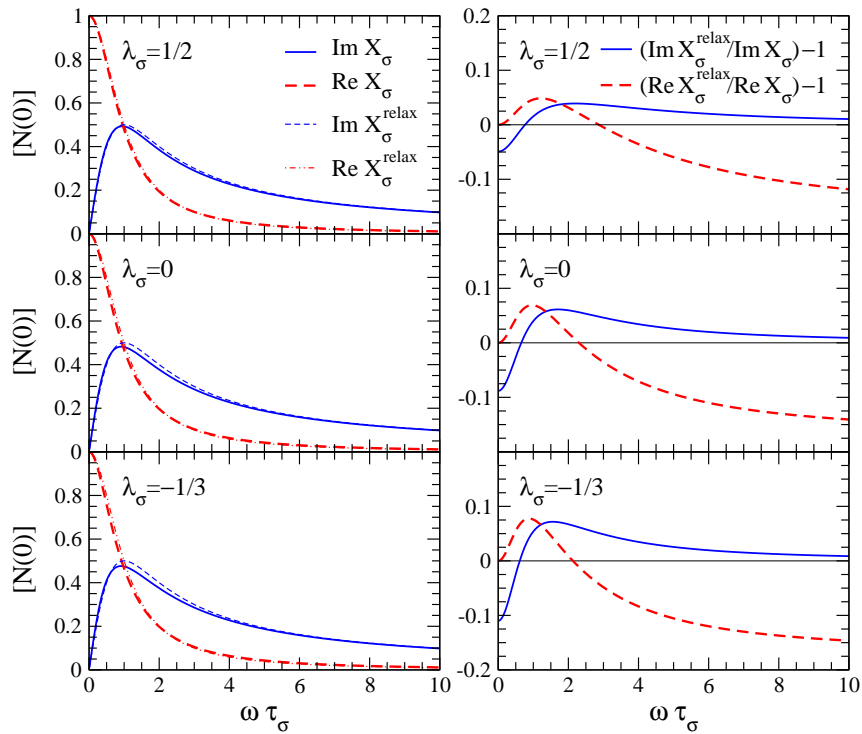}
\end{center}
\caption{(Color online) Left panels: Real and imaginary parts of the 
response function $X_\sigma$ in units of the density of states at the
Fermi surface $N(0)$, for $\lambda_\sigma=1/2, 0$ and $-1/3$, as a
function of $\omega \tau_\sigma$. The spin relaxation time
$\tau_\sigma$ is given by Eq.~(\ref{tau_sigma}). For comparison, we show
the response function based on the relaxation time approximation 
$X_\sigma^{\rm relax}$. Right panels: Relative
deviations of the real and imaginary parts of the relaxation time
approximation and the exact result, $({\rm Re} X_\sigma^{\rm
relax}/{\rm Re} X_\sigma)-1$ and $({\rm Im} X_\sigma^{\rm
relax}/{\rm Im} X_\sigma)-1$, for the same values of
$\lambda_\sigma$.\label{fig:response+ratio}} 
\end{figure*}

In Ref.~\cite{lykasov} a single relaxation time approximation was
made for the collision term, and this led to the response function
(in the absence of mean-field effects)
\be
X_\sigma^{\rm relax} = \frac{N(0)}{1 - \rmi \omega \tau_\sigma} \,.
\label{chirelax}
\ee
In terms of $\tau_\sigma$, the quantity $\zeta$ is given by
\be
\zeta^2 = 1 - \rmi \, \frac{4 \omega \tau_\sigma}{3} \,
(1-\lambda_\sigma) \,,
\ee
and this ensures that the relaxation time approximation,
Eq.~(\ref{chirelax}), reproduces the leading term $\rmi N(0)/(\omega
\tau_\sigma)$ in the response function for $\omega \tau_\sigma \gg 1$.
This can be shown by observing that, in this limit, the leading 
contribution to the deviation function is given by $\xi^{(0)}(x)
= 1/[2 \cosh(x/2)]$. When $\omega \tau_0$ is large, the leading
contributions to $\xi^{(1)}$ due to collisions can be calculated
in perturbation theory, treating $1/(\omega \tau_0)$ as small
parameter, and one finds
\begin{align}
\xi^{(1)}(x_1) &= - \frac{\rmi}{2\omega \tau_0} \biggl( (x_1^2+\pi^2) \,
\xi^{(0)}(x_1) \nonumber \\[1mm]
&\quad - \lambda_\sigma \int_{-\infty}^\infty \frac{dx_2 dx_3 dx_4 \,
\delta_x}{4 \cosh(x_2/2) \cosh(x_4/2)} \, \xi^{(0)}(x_3) \biggr)
\nonumber \\[1mm]
&= - \frac{\rmi}{2\omega \tau_0} \, (x_1^2 + \pi^2) \,
(1-\lambda_\sigma) \, \xi^{(0)}(x_1) \,,
\end{align}
where we have used the fact that $1/[2 \cosh(x/2)]$ is the
eigenfunction of the integral operator with eigenvalue 1, see
Eq.~(\ref{Q0}). From this, one finds that leading contribution
to the response function is given by $X_\sigma=\rmi N(0)/(\omega
\tau_\sigma)$ at high frequency. The analogous result for
spin-independent distortions of the Fermi surface was derived
in Ref.~\cite{pethick}.

\section{Examples}
\label{examples}

One case that can be solved simply is that for $\lambda_\sigma=0$, 
and Eq.~(\ref{Kxi}) then gives
\be
\xi(x) = - \frac{\rm i \omega \tau_0}{(x^2 + \pi^2 \zeta^2)
\cosh(x/2)} \,.
\ee
Substituting this expression into Eq.~(\ref{chi0}) leads to the
response function~\cite{3He}
\be
X_\sigma = N(0) \biggl(1 + \frac{\rmi \omega \tau_0}{\pi^2 \zeta} \,
\psi'\bigl((1+\zeta)/2\bigr) \biggr) \,,
\label{Xsigma0}
\ee
where $\psi'(z)$ is the trigamma function~\cite{abramowitz}. In this
limit, numerical results using Eq.~(\ref{Xsigma0}) agree very well
with those obtained from the general expression,
Eq.~(\ref{chi0final}), thereby providing a check on the latter.

In the middle-left panel of Fig.~\ref{fig:response+ratio}, we show the
real and imaginary parts of the response function $X_\sigma$ for
$\lambda_\sigma=0$ as a function of $\omega \tau_\sigma$. For
comparison, results based on the relaxation time approximation,
Eq.~(\ref{chirelax}), are also given. The physically relevant time
scale that determines the transition between collisionless and
hydrodynamic behavior is $\tau_\sigma$ which, according to
Eq.~(\ref{tau_sigma}), is shorter than $\tau_0$ by a factor
$3/(2\pi^2) \approx 0.15$ for $\lambda_\sigma=0$. In the right panel
of Fig.~\ref{fig:response+ratio}, the relative differences between the
relaxation time approximation and the exact result are shown more
clearly, by plotting $({\rm Re} X_\sigma^{\rm relax}/{\rm Re}
X_\sigma)-1$ and $({\rm Im} X_\sigma^{\rm relax}/{\rm Im}
X_\sigma)-1$. In the top and bottom panels of
Fig.~\ref{fig:response+ratio}, we present results for two other values
of $\lambda_\sigma$, relevant for neutrino interactions and the spin
response of neutron matter. These are $\lambda_\sigma=1/2$, which
corresponds to a ratio of the quasiparticle relaxation time to the
spin relaxation time, $\tau/\tau_\sigma = 3 \tau_0/(2 \pi^2
\tau_\sigma) = 1/2$, typical for low-momentum two-nucleon interactions
(see Fig.~3 in Ref.~\cite{lykasov}) with very similar rates obtained
in chiral effective field theory~\cite{bacca}, and
$\lambda_\sigma=-1/3$, which corresponds to $\tau/\tau_\sigma =4/3$,
the value for the one-pion exchange approximation to nuclear
interactions~\cite{lykasov}.

\begin{figure}[t]
\begin{center}
\includegraphics[clip=,scale=0.46]{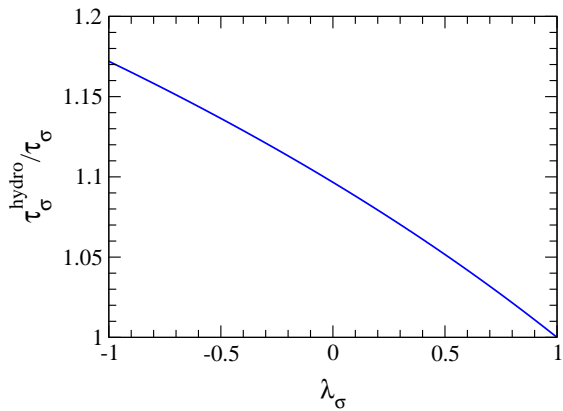}
\end{center}
\caption{Ratio of spin relaxation times in the hydrodynamic and
collisionless limits $\tau_\sigma^{\rm hydro}/\tau_\sigma$ as a
function of $\lambda_\sigma$.\label{fig:ratiooftaus}}
\end{figure}

At low frequencies the relaxation time approximation always
underestimates ${\rm Im} X_\sigma$. This may be understood in terms of
the standard variational calculation of the relaxation time in the
hydrodynamic limit. Generally, one can write the response
function at low frequencies as
\be
X_\sigma \approx N(0) \bigl( 1+\rmi \omega \tau_\sigma^{\rm hydro} \bigr) \,,
\label{chihydro}
\ee
where $\tau_\sigma^{\rm hydro}$ is a characteristic relaxation time in
this limit. According to Eqs.~(\ref{chi0final}) and~(\ref{chihydro}),
the ratio of the spin relaxation times in the hydrodynamic and 
collisionless limits is given by
\begin{align}
\frac{{\tau_\sigma}^{\rm hydro}}{\tau_\sigma} &=
\frac{4(1-\lambda_\sigma)}{3}
\nonumber \\
&\times \sum_{r=0}^\infty \, \frac{4r+3}{(2r+1)(2r+2)
[(2r+1)(2r+2)-2\lambda_\sigma]} \,.
\label{ratiooftaus}
\end{align}
This result may also be obtained directly by generalizing the
calculations of Refs.~\cite{jensen, bs} to the problem of spin
relaxation in a system with noncentral interactions.  The quantity
$\tau_\sigma$ is the simplest variational estimate of the
relaxation time and consequently always underestimates the relaxation
time $\tau_\sigma^{\rm hydro}$ in this limit. In
Fig.~\ref{fig:ratiooftaus}, we show the ratio $\tau_\sigma^{\rm
hydro}/\tau_\sigma$ as a function of
$\lambda_\sigma$. For systems, in which collisions only occur between particles
with the same spin projection and always change the sign of the spin
projection, one has $\lambda_\sigma = -1$ and the lack of spin conservation 
is maximal. In this case, ${\tau_\sigma}^{\rm hydro}/\tau_\sigma
\approx 1.17$, while for $\lambda_\sigma=0$ the ratio is 
${\tau_\sigma}^{\rm hydro}/\tau_\sigma = \pi^2/9 \approx 1.10$ and for
$\lambda_\sigma \to 1$ one finds ${\tau_\sigma}^{\rm hydro}/\tau_\sigma \to 1$.

The relaxation time approximation overestimates ${\rm Im} X_\sigma$
at higher frequencies.  That the differences of the imaginary parts
cannot have the same sign for low and high frequencies is a necessary
consequence of the fact that the response functions satisfy a 
Kramers-Kronig relation, and therefore
\begin{multline}
\frac{2}{\pi} \, \int_0^{\infty} \, \frac{d\omega}{\omega} \bigl[
{\rm Im} X_\sigma(\omega) - {\rm Im} X_\sigma^{\rm relax}(\omega)
\bigr] \\[1mm]
= {\rm Re} X_\sigma(0) - {\rm Re} X_\sigma^{\rm relax}(0) = 0 \,.
\end{multline}
In addition, at high frequencies ${\rm Re} X_\sigma$ is always greater
than the relaxation time approximation. This is due to the fact that
the leading term at high frequencies is proportional to the average of
the square of the collision operator, while in the relaxation time
approximation this is replaced by the square of the average. For 
$\lambda_\sigma=0$, this ratio can be calculated simply, and one finds
\be
\frac{{\rm Re} X_\sigma^{\rm relax}}{{\rm Re} X_\sigma}
\to \frac{\biggl( \int_{-\infty}^\infty dx \, (x^2+\pi^2)
{\rm sech}^2(x/2) \biggr)^2}{\int_{-\infty}^\infty dx \, (x^2+\pi^2)^2
{\rm sech}^2(x/2)} = \frac{5}{6} \,,
\ee
in the limit $\omega \tau_\sigma \gg 1$.

In summary, our results show that the differences between the real and
imaginary parts of the spin response function calculated in the
relaxation time approximation and the exact result are generally less
than $10\%$ for the conditions relevant to neutrino interactions in
stellar collapse, supernovae and neutron stars. We do not expect
this conclusion to change when mean-field effects are included,
since for the densities of interest, the Landau
parameter $G_0 \approx 0.8$ is positive and not large (see, for
example, Ref.~\cite{RGnm}).
Consequently, the errors of using the relaxation time approximation
are small compared to the theoretical uncertainties in determining
the spin relaxation times in dense matter from nuclear forces.

\section{Discussion}

We first consider applications of the spin response function to
calculations of neutrino-pair bremsstrahlung and absorption and to
neutrino scattering. As shown in Ref.~\cite{Raffelt3}, if one neglects
the momentum transfer to the nuclear medium and Pauli blocking
effects for neutrinos, the rates of neutrino processes when averaged
over a Boltzmann distribution for the neutrinos are proportional to
integrals of the form
\be
I_\nu = \int_0^\infty d\omega \, \omega^\nu \,
\frac{{\rm Im} \chi_\sigma(\omega)}{1-\rme^{-\omega/T}} \,,
\ee
where the exponent ranges from $\nu = 0$ for neutrino scattering to
$\nu = 7$ for the rate at which neutrino-pair emission and absorption
equilibrate the temperature of the nucleons with the temperature of
the neutrinos and antineutrinos. For $\nu=0$ and provided the width
$(1+G_0)/\tau_\sigma$ is small compared with $T$ (which is a necessary
condition for the picture of long-lived quasiparticle excitations to
be applicable, see also Ref.~\cite{lykasov}),
one can expand the denominator to first order in $\omega/T$ and the
rate for neutrino scattering is then proportional to 
\be
\int_0^\infty d\omega \, \frac{{\rm Im} \chi_\sigma(\omega)}{\omega}
\sim \chi_\sigma(0) \,,
\ee
where the latter expression is a consequence of the Kramers-Kronig
relation. Therefore, for this case the differences between the
frequency dependences of $\chi_\sigma^{\rm relax}$ and $\chi_\sigma$
have no effect. For larger values of $\nu$, higher frequencies are
weighted more heavily than lower ones, and consequently the relaxation
time approximation will overestimate rates. However, the error can
never be greater than the maximum deviation of ${\rm Im}
\chi_\sigma^{\rm relax}$ from ${\rm Im} \chi_\sigma$.
 
There are a number of directions in which this work could be extended.
One is to consider the case when $\omega \gtrsim T$, and another is to
treat partially degenerate or nondegenerate nucleons.

\section{Concluding remarks}

In this paper, we have solved the quasiparticle transport equation for
the spin response of a normal Fermi liquid for arbitrary values of
$\omega \tau_\sigma$. In a field-theoretical framework, this amounts
to solving the Bethe-Salpeter equation for a quasiparticle-quasihole
pair excitation, taking into account self-energy and vertex
corrections at the level of two-body collisions. One of the advantages
of the transport equation approach is that conservation laws are
properly taken into account.

In particular, we have studied the relaxation of the isotropic part
of the quasiparticle distribution function in momentum space. However,
the basic formalism can be applied to relaxation of other spherical
harmonic components, with appropriate changes in the definition of the
quantity corresponding to $\lambda_\sigma$. These methods may also be
applied to multicomponent systems, the main difference being that the
coefficients $a_r$ then depend on the species, and one must solve a
set of linear equations. The techniques therefore represent a
generalization of those used to calculate transport coefficients of
multicomponent Fermi liquids~\cite{quader}.
 
The differences between the exact solution and the relaxation time
approximation are small compared with the uncertainty of other
input to calculations of neutrino processes in dense matter. However,
the effects could be of importance in related situations, such as the
absorption of sound in liquid $^3$He for conditions between the
collisionless and hydrodynamic regimes. Another potential application
is to calculate the frequency-dependent conductivity of a charged
Fermi liquid and the magnetoresistance of metals, problems attacked in
Ref.~\cite{Ah-Sam} by variational techniques.

\section*{Acknowledgements}

We are grateful to Henrik Smith for helpful conversations and to John
Avery for introducing us to the uses of Sturmian methods in other
areas. We also thank Peter Orland and Jakob Yngvason for useful
remarks. AS thanks the Niels Bohr International Academy and NORDITA
for the warm hospitality. This work was supported in part by the
Natural Sciences and Engineering Research Council of Canada (NSERC).
TRIUMF receives funding via a contribution through the National
Research Council Canada.

\appendix

\section{Mathematical details}
\label{math}

To make the paper reasonably self-contained, we list a number of
useful mathematical relations in this appendix.  In the derivation of
the quasiparticle transport equation, Eq.~(\ref{Kxi}), we encountered
the integrals
\be
K_2(x) = \int_{-\infty}^\infty \frac{dx_2 dx_4 \,
\delta(x+x_2-x_4)}{4 \cosh(x_2/2) \cosh(x_4/2)} \,,
\ee
and
\be
K_3(x) = \int_{-\infty}^\infty \frac{dx_2 dx_3 dx_4 \,
\delta(x+x_2-x_3-x_4)}{8 \cosh(x_2/2) \cosh(x_3/2)
\cosh(x_4/2)} \,.
\ee
These may be evaluated by representing the delta function as
\be
\delta(x) = \int_{-\infty}^\infty \, \frac{dt}{2\pi} \: \rme^{\rmi t x} \,,
\label{delta}
\ee
and using the methods described in Appendix~A of Chapter~1 of
Ref.~\cite{BaymPethick}. The results are
\be 
K_2(x) = \frac{x}{2 \sinh(x/2)} \,,
\label{K2}
\ee
and 
\be
K_3(x) = \frac{x^2+\pi^2}{2} \, \frac1{2 \cosh(x/2)} \,.
\label{K3}
\ee

By multiplying Eq.~(\ref{Phi_r}) by $\widetilde{\Phi}_{r'}(k)$, subtracting
the same expression with $r$ and $r'$ interchanged, and then integrating
over $k$, one sees that the $\widetilde{\Phi}_r(k)$ are orthogonal in
the sense that
\be
\int_{-\infty}^\infty dk \: \frac{\widetilde{\Phi}_r(k) \, \widetilde{
\Phi}_{r'}(k)}{\cosh^2(\pi k)} = \delta_{r r'} \, \frac{U_r}{2\pi} \,,
\ee
where $U_r$ is given explicitly in Eq.~(\ref{Ur}).
For the eigenfunctions in terms of $x$, this condition is equivalent
to
\be
\int_{-\infty}^\infty dx \, (x^2 + \pi^2 \zeta^2) \, \Phi_r(x) \,
\Phi_{r'}(x) = \delta_{r r'} \, \frac{U_r \Lambda_r}{2} \,.
\ee

In the hydrodynamic limit ($\zeta\rightarrow 1$), the eigenfunctions
$\Phi_r(x)$ have a simple form, since
\begin{align}
\Phi_r(x) &= \int_{-\infty}^\infty \frac{dk}{2\pi} \:
\widetilde{\Phi}_r(k) \, \rme^{\rmi k x} \,, \nonumber \\[1mm]
&\sim \int_{-\infty}^\infty \frac{dk}{2\pi} \: {\rm sech}(\pi k) \,
C_{2r}^{3/2}\bigl(\tanh(\pi k)\bigr) \, \rme^{\rmi kx} \,,
\end{align}
where we have used Eq.~(\ref{gegenbauer}) for $\widetilde{\Phi}_r(k)$.
The integral can be evaluated by contour integration, closing the
contour in the upper half plane for $x>0$ and in the lower half plane
for $x<0$.  The integrand has poles of order $2r+1$ at $k=\rmi
(\nu+1/2)$ with integer $\nu$, and therefore the integral is
proportional to the sum of the residues of the integrand at these
points, which is a geometrical progression with the ratio of
subsequent terms equal to $-\rme^{-|x|}$.  The final result is
$\Phi_r(x) \sim Q_r(x)/[2 \cosh(x/2)]$, where $Q_r(x)$ is the residue
of $\rme^{\rmi k x} \, C_{2r}^{3/2}( \coth(\pi k) )/
\sinh(\pi k)$ at $k=0$, a polynomial of order $2r$. The polynomials
$Q_r(x)$ satisfy the orthogonality relation
\be
\int_{-\infty}^\infty
dx \: \frac{x^2+\pi^2}{\cosh^2(x/2)} \, Q_r(x) \, Q_{r'}(x) = 0 \,,
\quad {\rm for} \quad r \neq r' \,.
\ee
The first few eigenfunctions are given by
\begin{align}
Q_0(x) &\sim 1 \,, \label{Q0} \\[1mm]
Q_1(x) &\sim \biggl( 1 - \frac{5}{3} \, \frac{x^2}{\pi^2} \biggr) 
\,, \quad \text{and} \\[1mm]
Q_2(x) &\sim \biggl( 1 -\frac{14}{5} \, \frac{x^2}{\pi^2}
+ \frac{7}{15} \, \frac{x^4}{\pi^4} \biggr) \,.
\end{align}
For other values of $\zeta$ the eigenfunctions do not have such a
simple form.


\begin{thebibliography}{99}
\bibitem{normalization} Note that the average defined in
Ref.~\cite[Eq.~1.2.88]{BaymPethick} is normalized to 2.
\bibitem{notation} The quantity $\tau_0$ in this paper
is denoted by $\tau$ in Ref.~\cite{BaymPethick}.
\bibitem{3He} Essentially the same result was obtained
in studies of zero sound in liquid $^3$He in 
Ref.~\cite[Eq.~(60)]{egilsson}.
\end{thebibliography}
\end{document}